\documentclass[conf]{new-aiaa}
\usepackage[utf8]{inputenc}

\usepackage{graphicx}
\usepackage{amsmath}
\usepackage[version=4]{mhchem}
\usepackage{siunitx}
\usepackage{longtable,tabularx}
\usepackage{subcaption}
\usepackage{siunitx}

\newcommand{\ud}{\mbox{\textrm d}} 

\title{Influence of Surface Roughness Geometry on Trailing Edge Wall Pressure Fluctuations and Noise}

\author{Fernanda L. dos Santos\footnote{PhD candidate, Engineering Fluid Dynamics, Department of Thermal Fluid Engineering, f.l.dossantos@utwente.nl}, Nikolaj A. P. Even \footnote{Bachelor's student, Mechanical Engineering, nikolaj.even@icloud.com}, Laura Botero-Bolívar \footnote{PhD candidate, Engineering Fluid Dynamics, Department of Thermal Fluid Engineering, l.boterobolivar@utwente.nl}, Cornelis H. Venner \footnote{Chair, Engineering Fluid Dynamics group, Department of Thermal Fluid Engineering, AIAA member, c.h.venner@utwente.nl} and Leandro de Santana, \footnote{Assistant Professor, Engineering Fluid Dynamics, Department of Thermal Fluid Engineering, AIAA member, leandro.desantana@utwente.nl}}
\affil{University of Twente, Drienerlolaan 5, 7522 NB, Netherlands}

\begin{document}

\maketitle

\begin{abstract}
Surface roughness elements are commonly used in wind tunnel testing to hasten the laminar-turbulent transition of the boundary layer in model tests to mimic the aerodynamic effects present in the full-scale application. These devices can alter the characteristics of the turbulent boundary layer, such as the spanwise correlation length, the boundary layer thickness, the displacement thickness, etc. This not only affects the aerodynamic performance but also the aeroacoustic characteristics of the tested model. Few studies have investigated the effects of the surface roughness elements on the trailing edge near- and far-field noise. So far, the influence of roughness on the wall pressure fluctuations and spanwise coherence at the trailing edge has been left unexplored. Thus, this research addresses the effects of surface roughness geometries of different heights on the trailing edge wall pressure fluctuations, the spanwise coherence, and the far-field noise. The experiments were performed in the Aeroacoustic Wind Tunnel of the University of Twente adopting zigzag strips and novel sharkskin-like surface roughness installed in a NACA 0012 at zero angle of attack. The tested surface roughness heights ranged from 29\% to 233\% of the undisturbed boundary layer thickness. The wall pressure fluctuations and the far-field noise were measured for Reynolds numbers from \boldmath{$1.3~\times~10^5\text{~to~}3.3~\times~10^5$}. It was observed that the surface roughness affects the low- and high-frequency range of the wall pressure spectrum, with trip heights in the range from 50\% to 110\% of the undisturbed boundary layer thickness having a slight level increase for low frequencies and no difference for high frequencies. The far-field noise increased for low and high frequencies as the trip height increased. The low-frequency increase is a consequence of the trip effects on the trailing edge wall pressure fluctuations, whereas for high frequencies the increase is due to the noise generated by the trip itself. Moreover, the sharkskin-like trip showed to be an effective tripping device. However, this geometry results in high levels of far-field noise in the high-frequency range.
\end{abstract}

\section{Nomenclature}

{\renewcommand\arraystretch{1.0}
\noindent\begin{longtable*}{@{}l @{\quad=\quad} l@{}}
$b_c$   & Corcos' constant [-]\\
$c$     & airfoil chord length [m]\\
$C_f$   & skin friction coefficient [-]\\
$d$     & airfoil span length [m]\\
$f$     & frequency [Hz]\\
$H$     & sharkskin-like surface roughness height [m], boundary layer shape parameter [-]\\
$H_k$   & boundary layer kinematic shape parameter [-]\\
$k$     & tripping device height [m]\\
$L$     & sharkskin-like surface roughness length [m]\\
$l_y$   & spanwise correlation length [m]\\
$M_e$   & boundary layer edge Mach number [-]\\
$P_{i,j}$ & cross-power spectral density [$\text{Pa}^2/\text{Hz}$]\\
$P_{i,i}$ & auto-power spectral density [$\text{Pa}^2/\text{Hz}$]\\
$P$\textsubscript{ref} & reference pressure, $P$\textsubscript{ref} = 20 $\upmu$Pa \\
$PSD$   & power spectrum density [$\text{Pa}^2/\text{Hz}$]\\
$Re_k$   & trip based Reynolds number [-]\\
$St$   & Strouhal number based on the boundary layer thickness and edge velocity [-]\\
$U$     & free-stream velocity [m/s]\\
$U$\textsubscript{c}   & convection velocity [m/s]\\
$U$\textsubscript{e}   & edge velocity at the trailing edge [m/s]\\
$u_k$   & downstream velocity at the trip height [m/s]\\
$u_{\tau}$ & friction velocity [m/s]\\ 
$W$     & sharkskin-like surface roughness length [m]\\
$x$     & downstream direction [m]\\
$y$     & spanwise direction [m]\\
$z$     & normal-to-the-surface direction [m]\\
$\gamma^2$ & coherence [-]\\
$\delta$ & boundary layer thickness [m]\\
$\delta_k$ & undisturbed boundary layer thickness at the trip position [m]\\
$\delta^*$ & boundary layer displacement thickness [m]\\
$\eta_y$    & spanwise distance between the microphones [m]\\
$\theta$ & boundary layer momentum thickness [m]\\
$\nu$ & kinematic viscosity [m\textsuperscript{2}/s]\\
$\phi_{i,j}$ & phase between two microphone signals [rad]\\
$\Phi_{pp}$ & Power spectral density of the wall pressure fluctuations [$\text{Pa}^2/\text{Hz}$]\\
\end{longtable*}}

\section{Introduction} \label{sec: introduction}

Aerodynamic noise pollution impacts people's well-being and animal life motivating the urgent need for the reduction of this noise source. The societal relevance of flow-induced noise has contributed to the intensification of this research topic~\cite{Regulation,Guidelines}. Noise generated by the interaction between the unsteadiness present in the turbulent boundary layer and the sharp corner of the trailing edge (TE) of a foil or other fluid dynamic devices~\cite{Devenport} is known as trailing edge noise. This aeroacoustic mechanism is a relevant noise source of wind turbines~\cite{OERLEMANS2007}, aircraft~\cite{Zhang2010}, and ships~\cite{Carlton}. The TE far-field noise intensity and spectrum depend mainly on the turbulent boundary layer characteristics at the trailing edge~\cite{Brooks1989}. Thus, describing the turbulent flow at the trailing edge is paramount to understanding and modelling trailing edge noise.

Wind tunnel tests are an essential tool to study the physics and the characteristics of turbulent boundary layers and trailing edge noise. However, some difficulties are intrinsic to these experiments. Wind tunnel measurements aim to mimic the operating conditions encountered in real-size applications. Due to cost and size limitations scaled models are used in the wind tunnel tests, where low free-stream velocities are commonly encountered. Consequently, the Reynolds number in these tests are usually one to two orders of magnitude smaller than for the full-scale application. This Reynolds number mismatching leads to difficulties in guaranteeing flow similarity between the scaled tests and the full-size application. This problem is commonly overcome by the use of surface roughness element, also referred to as tripping device, or trip, placed at a strategic position on the surface of the scaled model~\cite{Rona_2010,Reshotko_1976,Fernanda2020}. 

%

Surface roughness elements hasten the boundary layer transition from laminar to turbulent flow. These elements are used extensively in wind tunnel tests for aerodynamic performance analysis and aeroacoustic measurements. Surface roughness also develops during the lifetime of a piece of equipment. Wind-turbine and gas-turbine blades erode due to particle impact, resulting in the formation of surface roughness close to the leading edge~\cite{Sareen2014,Hamed2004}. Previous research has shown that the type of surface roughness and its height have a direct influence on the transition onset and the turbulent boundary layer development~\cite{Fernanda2020,Silvestri_2018,Elsinga2012,Ye2021,Chan_1988,Braslow_1966}, consequently also affecting the trailing edge noise. Braslow~\cite{Braslow_1966} stated that for two-dimensional surface roughness, e.g., zigzag strips, spots of turbulence begin to move forward of the natural position of transition when a critical value of Reynolds number based on the trip height is reached. A further increase in Reynolds number, i.e., trip height, is required to move the transition gradually forward. Therefore, a minimum height for the surface roughness exists to result in a faster transition. Dos Santos et al.~\cite{Fernanda2020} investigated the development of a turbulent boundary layer on a flat plate by means of zigzag and grit surface roughness. They observed that the zigzag strip did not trigger transition directly behind it and also the turbulent flow developed for different roughness heights did not resemble a naturally developed turbulent flow in the viscous sublayer and the buffer layer, not even at distances as far as 300~$\delta_k$ downstream of the roughness. Elsinga et al.~\cite{Elsinga2012} performed tomographic PIV measurements of the flow developed by a zigzag strip, visualizing and evaluating flow structures in a number of downstream positions. They observed individual turbulent structures in the flow downstream of the trip that are typical of a developed turbulent boundary layer, e.g., hairpins and packets. However, there was also a large spanwise coherence in the flow, which they attributed to the surface roughness elements. Ye~\cite{Ye2021} measured the transition point of the boundary layer by means of infrared thermography, observing that the transition location for a zigzag strip of 0.3~mm high moves closer to the strip as the free-stream velocity increases. This occurs because for higher velocities, the boundary layer becomes thinner, making the roughness height more comparable with the boundary layer thickness. The surface roughness thus plays an important role in the transition and its onset, and in the characteristics of the turbulent flow. Therefore, choosing an appropriate roughness geometry and height is crucial for accurate measurement of the turbulent boundary layer, and consequently the far-field noise.

\begin{figure}[b!]
	\centering
    \includegraphics[width=0.49\textwidth]{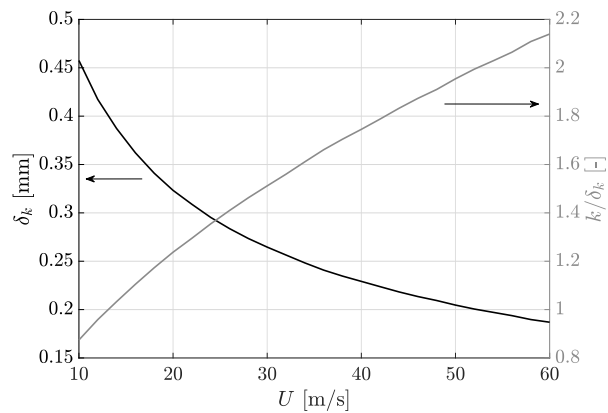}
	\caption{Undisturbed boundary layer thickness (\boldmath{$\delta_k$}) and ratio between the roughness height ($k$) and \boldmath{$\delta_k$} for a velocity sweep. The boundary layer thickness was obtained from XFOIL at \boldmath{$x/c=0.05$}, considering natural transition developed in a NACA 0012 with a chord length of \boldmath{$200$}~mm. Reynolds number ranging from~\boldmath{$1.3~\times~10^5\text{~to~}8~\times~10^5$}. The roughness height considered was \boldmath{$0.4$}~mm, which is commonly applied in small-scale wind tunnel tests.}
	\label{fig:trip_height}
\end{figure}

A critical parameter in determining which roughness height ($k$) to use is its ratio with the undisturbed boundary layer thickness at the trip location ($\delta_k$)~\cite{Fernanda2020,Silvestri_2018}. When $k/\delta_k$ is too small it does not introduce enough instabilities in the boundary layer to develop a turbulent flow at the TE~\cite{Ye2021} and a too high ratio changes the boundary layer characteristics~\cite{Fernanda2020,Silvestri_2018}. A common practice in wind tunnel tests is to perform velocity sweeps with the same roughness height, as performed by Ye et al.~\cite{Ye2021}. In these cases, the boundary layer becomes thinner as the velocity increases, with the roughness height becoming much higher than the boundary layer thickness, see Fig.~\ref{fig:trip_height}. If this effect is not taken into consideration when performing aeroacoustic and aerodynamic measurements, considerable errors can be made in determining the turbulent boundary layer parameters and the wall pressure fluctuations (WPF), also refereed as unsteady surface pressure. This becomes an even greater concern when experimental data is used to not only validate WPF models but also to conceive them since they are widely used as an input for TE noise prediction models. 
Therefore, determining the appropriate roughness height is of great importance to perform aeroacoustic and aerodynamic measurements.

As the surface roughness elements affect the turbulent boundary layer characteristics, the TE noise is also affected by the roughness geometry and height. So far, little research has been carried out to study the roughness effects on the TE noise. Hutcheson and Brooks~\cite{Hutcheson} measured TE noise from a NACA 63-215 using an aluminium serrated strip and steel grits over the first 5\% of the chord. It was found that the noise source close to the leading edge due to the surface roughness was dominant for frequencies higher than 12.5~kHz. Cheng et al.~\cite{Cheng2018} studied the effects of ice-induced surface roughness on rotor broadband noise. They observed a considerable increase in the trailing edge noise for frequencies higher than 4~kHz from far-field measurements. They associated this noise increase to an increase of the boundary layer turbulence intensity at the trailing edge due to the roughness, which was observed from numerical simulations. Ye et al.~\cite{Ye2021} performed far-field measurements on a NACA~0012 with grits and zigzag strip of 0.3~mm height for velocities from 16~m/s to 35~m/s. They observed that the far-field noise levels for different velocities were closely related to the tripping conditions and to the unsteady flow features developed in the transition process. Therefore, surface roughness affects the turbulent boundary layer at the TE and consequently the TE far-field noise. However, the effects of the roughness geometry and height on the wall pressure fluctuations at the TE has not been investigated before, even though the wall pressure is one of the most important parameters for TE noise modeling.

In this study, the effects of two surface roughness geometries on the wall pressure fluctuations and the far-field noise at the trailing edge are investigated. The wall pressure was measured by microphones installed on a NACA~0012 airfoil close to the TE. The far-field noise was measured by a free-field microphone perpendicular to the airfoil TE. The roughness elements used in this research are a zigzag strip and a new sharkskin-like trip. The zigzag strip was chosen since it is extensively applied in wind tunnel testing, whereas the novel sharkskin-like trip was chosen since its geometry has lower aerodynamic drag~\cite{Domel2018}. Consequently, it is expected to have a smaller effect on the turbulent boundary layer. Different heights of the surface roughness were tested.

\section{Experimental setup and methodology}
\subsection{Wind tunnel}
The experiments were performed in the UTwente AeroAcoustic Wind Tunnel which is an aeroacoustic open-jet, closed-circuit wind tunnel. The test section dimensions are 0.7~m $\times$ 0.9~m and it has a contraction ratio of 10:1. After the contraction, the flow enters a closed test section (CTS) and subsequently an open test section (OTS). The maximum operating velocity is 60~m/s in the open-jet configuration with a turbulence intensity below 0.08\% \cite{Leandro_2018}. An anechoic chamber of 6 $\times$ 6 $\times$ 4~m\textsuperscript{3} with a 160~Hz cut-off frequency encloses the test region. The flow temperature was controlled and maintained at approximately 20~°C for all experiments. Experiments were performed at four free-stream velocities ($U$): 10, 15, 20, and 25~m/s.
\begin{figure}[b!]
	\centering
    \begin{subfigure}[c]{0.55\textwidth}
        \centering
		\includegraphics[width=\textwidth]{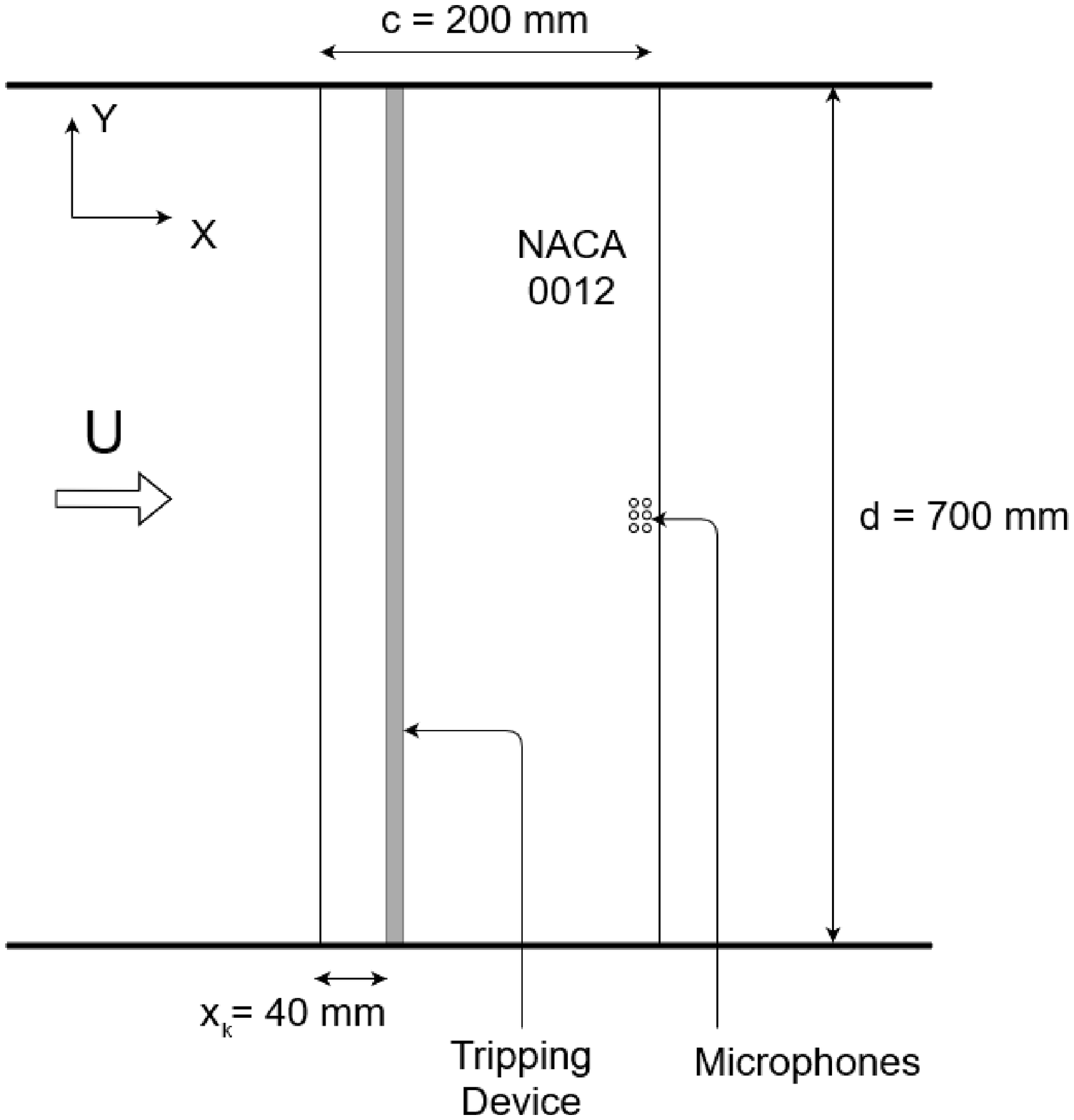}
    \end{subfigure}
	\hfill
	\begin{subfigure}[c]{0.35\textwidth}
        \centering
        \includegraphics[width=\textwidth]{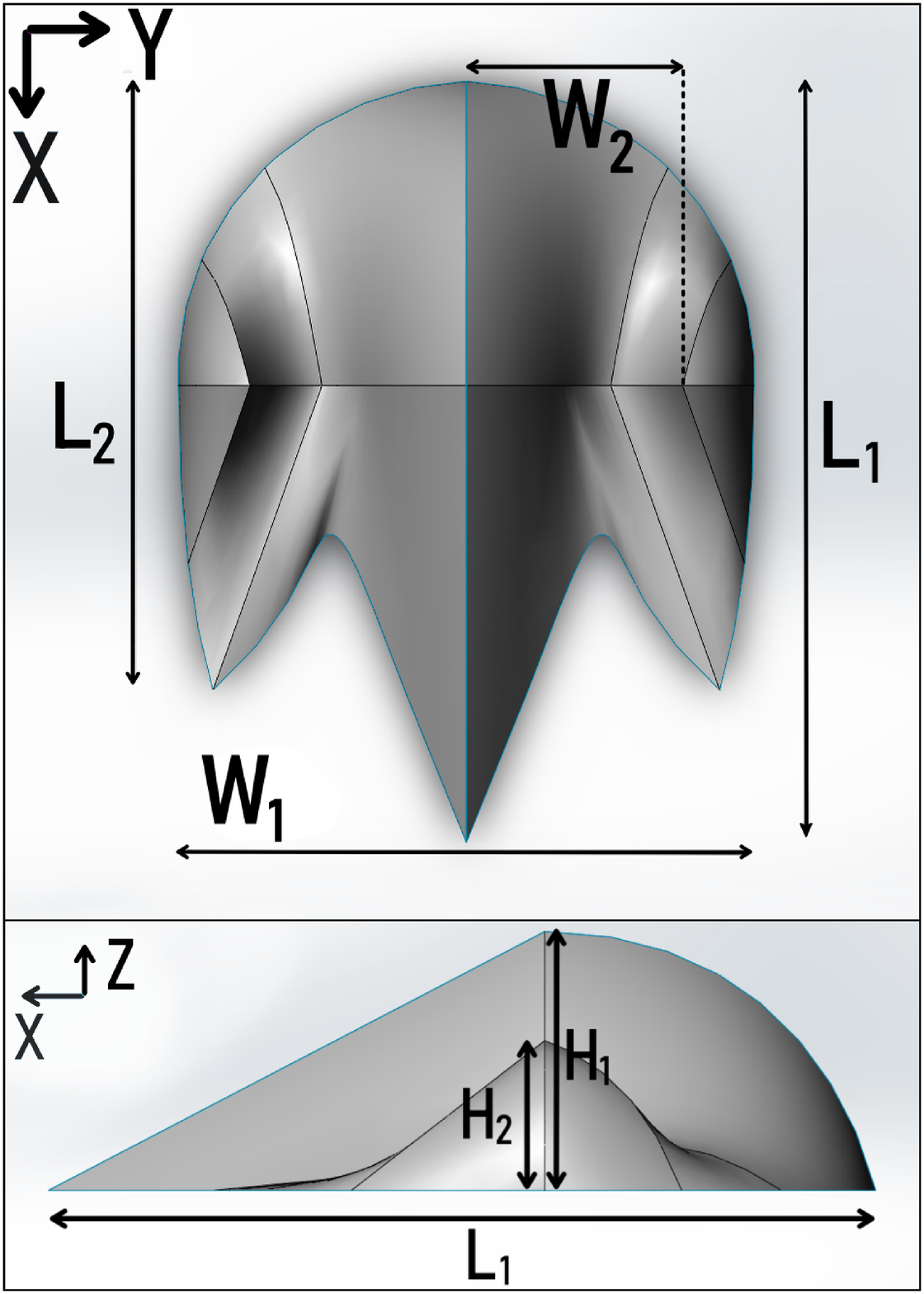}
    \end{subfigure}
	\caption{Experimental setup (left) and sharkskin-like geometry for the tripping device (right). The coordinate systems shown in the figures are the same, so that the orientation of the sharkskin on the airfoil surface can be inferred. Details of the geometries tested are described in section \ref{sec:surface_roughness}.}
	\label{fig:exp_setup_sharskin_geom}
\end{figure}

\subsection{Airfoil model and instrumentation}
A NACA~0012 with chord and span lengths of $c=200$~mm and $d=700$~mm was used. The angle of attack was maintained at $0^\circ$ in all cases. The Reynolds numbers based on the chord length ranged from $1.3~\times~10^5\text{~to~}3.3~\times~10^5$. The surface roughness was placed at $x/c=0.2$. The airfoil was instrumented with FG-23329-P07 Knowles microphones under pinholes of 0.3~mm near the TE (see Fig.~\ref{fig:exp_setup_sharskin_geom}). Six microphones were installed in two rows of 3 spanwise distributed sensors. One row of microphones is located at $x/c = 0.9275$ and the other at $x/c = 0.9125$, defining $x = 0$ at the leading edge (LE). The two central chordwise microphones are located at the midspan, with the other microphones spaced by 3~mm in the spanwise direction. The data was acquired at a sampling frequency of approximately 65.5~kHz ($2^{16}$~Hz) using PXIe-4499 Sound and Vibration modules embedded in a NI~PXIe-1073 chassis. The acquisition time was 30 seconds. 

\subsubsection{Calibration}
The microphones under a pinhole were calibrated to determine their sensitivity and frequency response. The sensitivity calibration was performed using an in-house calibrator with a GRAS~40PH free-field microphone as reference and a 1~kHz tone as the noise source. The frequency response was also determined using an in-house calibrator with a GRAS~40PH free-field microphone as reference but with white noise as the noise source. The sensitivity and the frequency calibrations were applied to all microphones.

\subsection{Boundary layer parameters}
The boundary layer parameters used in this study were obtained from XFOIL simulations~\cite{Drela_XFOIL}. The boundary layer displacement thickness ($\delta^*$), momentum thickness ($\theta$), skin friction coefficient ($C_f$) and edge velocity ($U_e$) were obtained from XFOIL for each free-stream velocity tested. From these parameters, the boundary layer thickness was calculated as~\cite{Drela}:
\begin{equation}
    \delta = \theta \left( 3.15 + \frac{1.72}{H_k-1}+\delta^* \right)\,,
\end{equation}
where $H_k$ is the kinematic shape parameter and it is given as~\cite{Drela}:
\begin{equation}\label{eq:kinematic_shape_parameter}
    H_k = \frac{H-0.290 M_e^2}{1+0.113 M_e^2} \,,
\end{equation}
where $M_e$ is the boundary layer edge Mach number, and $H$ is the shape parameter given as $H=\delta^*/\theta$. For low mach number flows, $H_k$ given by Eq.~\ref{eq:kinematic_shape_parameter} reduces to $H$. This approximation was used in this study. The undisturbed boundary layer thickness $\delta_k$ was determined from XFOIL considering natural transition, where the natural transition occurred downstream of the trip location. For the boundary layer parameters at the trailing edge, the transition was set at $x/c=0.2$.

The friction velocity $u_\tau$ was determined from the skin friction coefficient as:
\begin{equation}
    u_\tau = \sqrt{ 0.5 U^2 C_f}\,,
\end{equation}
and used to make the wall pressure fluctuations nondimensional. It is important to highlight that the boundary layer parameters obtained from XFOIL do not depend on the roughness geometry and height. Hence, the boundary layer parameters at the TE are dependent only on the free-stream velocity, with different roughness height and geometry presenting constant boundary layer values at the TE. Table~\ref{tab:trip} shows the boundary layer parameters at the trip position and at the TE obtained from XFOIL simulations for the free-stream velocities tested.
\begin{table}[t!]
\centering
\caption{Undisturbed boundary layer thickness (\boldmath{$\delta_k$}) at the trip position (\boldmath{$x/c=0.2$}) and boundary layer parameters used in the wall pressure spectrum nondimensionalization at the trailing edge for the free-stream velocity tested. The boundary layer parameters were obtained from XFOIL for a NACA 0012 with a chord length of \boldmath{$200$}~mm. Trip heights (\boldmath{$k$}) tested and the ratio between the trip height and the undisturbed boundary layer thickness at the trip location.}
\label{tab:trip}
\begin{tabular}{cclccclccccc}
\hline
       &  At trip position &  & \multicolumn{3}{c}{At trailing edge} &  & \multicolumn{5}{c}{$k$ {[}mm{]}}          \\ \cline{8-12} 
       & $x/c=0.2$ &  & \multicolumn{3}{c}{$x/c=0.95$}                       &  & 0.3    & 0.5    & 0.8    & 1.2    & 1.5   \\ \cline{2-2} \cline{4-6} \cline{8-12} 
       & $\delta_k$ {[}mm{]}          &  & $\delta$ {[}mm{]} & $U_e$ {[}m/s{]} & $u_{\tau}$ {[}m/s{]} &  & \multicolumn{5}{c}{$k/\delta_k$ {[}\%{]}} \\ \cline{2-2} \cline{4-6} \cline{8-12} 
10 m/s & 1.02                         &  & 9.3               & 9.16            & 0.33                 &  & 29     & 49     & 79     & 118    & 147   \\
15 m/s & 0.83                         &  & 8.4               & 13.7            & 0.48                 &  & 36     & 60     & 96     & 144    & 180   \\
20 m/s & 0.72                         &  & 7.9               & 18.2            & 0.63                 &  & 42     & 69     & 111    & 167    & 208   \\
25 m/s & 0.64                         &  & 7.5               & 22.7            & 0.77                 &  & 47     & 78     & 124    & 186    & 233   \\ \hline
\end{tabular}
\end{table}

\subsection{Surface roughness} \label{sec:surface_roughness}
Two types of surface roughness elements were investigated in this research: zigzag strips and sharkskin-like trips. Zigzag strips were chosen because they are widely used in wind tunnel tests due to their ease of installation and effectiveness in hastening the boundary layer transition~\cite{Silvestri_2018,Fernanda2020}. In this study, zigzag strips of $60^{\circ}$ teeth angle, 12~mm width and manufactured by Glasfaser-Flugzeug-Service GmbH were utilized. The other geometry tested is a sharkskin-like geometry. This geometry is considered as a possible tripping device due to its aerodynamic characteristics~\cite{Domel2018}. Domel et al.~\cite{Domel2018} investigated the aerodynamic properties of sharkskin-like geometry at low Reynolds number, observing that sharkskin-inspired trip improved the lift/drag ratio for airfoils. Furthermore, the authors noted that the sharkskin induced a small reattaching separation bubble in the boundary layer, which caused the formation of downstream vortices. Because of these properties, a sharkskin-like geometry seems to be a compelling choice for a tripping device; therefore, it is also investigated in this study.

\subsubsection{Sharkskin-like trip geometry and manufacturing process}
The geometry proposed by Domel et al.~\cite{Domel2018} was the source of inspiration to design the trip for this study. The sharkskin-like geometry investigated in this research was designed by scrutinizing microscopic images of actual sharkskin~\cite{Wen2014} to obtain a more organic geometry, see Fig.~\ref{fig:exp_setup_sharskin_geom}. The ratios between the indicated lengths are defined as follows: $L_2/L_1=0.8$, $W_2/W_1=0.38$, $W_1/L_1=0.76$, $H_2/H_1=0.58$, and $H_1/L_1=0.31$. A substrate of 0.6~mm was used to support the sharkskin elements. The trip height $k$ for the sharkskin is $H_1$ plus the substrate height. The sharkskin-like elements were distributed in rows where they were side by side in the spanwise direction. For subsequent rows, the sharkskin elements were placed in a staggered arrangement. The sharkskin-like trip was 12~mm wide, as the zigzag strip.


The sharkskin-like trip was 3D-printed at the Rapid Prototyping Lab of the University of Twente. The printing technique applied was Selective Laser Sintering. This technique uses a powder as input, and a laser is used to heat up the powder such that it hardens in the desired shape, allowing the unmelted powder to be simply blown off. The Selective Laser Sintering technique was chosen because, among the available techniques at the Rapid Prototyping Lab, it was the only one that could print the smallest trip wall thickness with details in the 10-100 micron range, and had low risk of warpage. The sharkskin trip was 3D-printed using Nylone (PA2200).

\subsubsection{Surface roughness height}
The maximum surface roughness height is determined based on the findings of~\cite{Silvestri_2018,Rengasamy_2017,Fernanda2020}. The maximum height is suggested to be equal to the undisturbed boundary layer thickness at the trip position $\delta_k$ in order to prevent overstimulation of the boundary layer, which leads to a non-canonical turbulent flow. However, it is quite common in wind tunnel tests to perform velocity sweeps without changing the roughness height, as discussed in Sec.~\ref{sec: introduction}. For high velocities this can result in a roughness height considerably higher than the undisturbed boundary layer thickness. Hence, the maximum roughness height tested was higher than the undisturbed boundary layer thickness, allowing the investigation of overtrip effects in the wall pressure fluctuations at the TE and far-field noise.

The minimum roughness height is dictated by its effectiveness in hastening the boundary layer transition, and it has been widely investigated in the literature~\cite{Braslow_1958,Fernanda2020,Silvestri_2018}. The most applied methodology to determine the minimum height is the one proposed by Braslow and Knox~\cite{Braslow_1958}. They defined the Reynolds number:
\begin{equation}
    Re_k = \frac{u_k k}{\nu}\,,
\end{equation}
where $u_k$ is the undisturbed boundary layer velocity at the roughness height, $k$ is the roughness height, and $\nu$ the fluid kinematic viscosity. Braslow and Knox~\cite{Braslow_1958} suggested that $Re_k \geq 600$ to obtain a fully developed turbulent boundary layer. Based on this Reynolds number, the minimum roughness height is: 0.97~mm (95\% $\delta_{k,10}$) for 10~m/s,  0.67~mm (81\% $\delta_{k,15}$) for 15~m/s, 0.53~mm (74\% $\delta_{k,20}$) for 20~m/s, and 0.44~mm (68\% $\delta_{k,25}$) for 25~m/s, where $\delta_{k,U}$ is the undisturbed boundary layer thickness at the trip location ($x/c=0.2$) for different free-stream velocities $U$. For this study, the minimum trip height was smaller than the minimum proposed by Braslow and Knox~\cite{Braslow_1958} in order to investigate the influence of roughness height in the development of a fully turbulent boundary layer.

The zigzag strip heights were chosen based on the heights available commercially. The minimum height applied was $k=0.3$~mm (29\% $\delta_{k,10}$ and 47\% $\delta_{k,25}$), which is smaller than the minimum height calculated by Braslow and Knox methodology for all wind tunnel speeds tested (95\% $\delta_{k,10}$ and 68\% $\delta_{k,25}$). The maximum height tested was $k=1.5$~mm (147\% $\delta_{k,10}$ and 233\% $\delta_{k,25}$). Intermediate heights were also tested to evaluate the effects of the roughness height on the wall pressure fluctuations and far-field noise in detail. The zigzag trip heights tested are presented in Table~\ref{tab:trip} with the boundary layer parameters at the TE obtained from XFOIL simulations with fixed transition at the trip-position.

The sharkskin-like trip height was determined based on the 3D-printing process limitations. The 3D-printing process restricted the minimum height possible, making it difficult to print sharkskin-like trip that was small relative to the boundary layer. Hence, the minimum height for the sharkskin-like trip was 1.5~mm, which corresponds to roughly 147\% of $\delta_{k,10}$ and 233\% of $\delta_{k,25}$.

\subsection{Wall pressure fluctuations}
\subsubsection{Spectrum}
The power spectral density (PSD) of the wall pressure fluctuations ($\Phi_{pp}$) is calculated based on Welch’s method. The data was averaged using blocks of $2^{13} = 8192$ (0.125~s) samples and windowed by a Hanning windowing function with 50\% overlap, resulting in a frequency resolution of 8~Hz. The PSD level of the WPF was nondimentionalized by the boundary layer parameters at the airfoil trailing edge: edge velocity $U_e$, friction velocity $u_\tau$, and boundary layer thickness $\delta$. The frequency ($f$) was nondimensionalized by the boundary layer parameters at the trailing edge as the Strouhal number:
\begin{equation} \label{eq:St}
    St = \frac{f \delta}{U_e}\,.
\end{equation}

\subsubsection{Spanwise coherence and convection velocity}
A crucial input parameter for TE noise predictions is the spanwise correlation length at the trailing edge, obtained from the coherence between two microphones in the spanwise direction. The coherence between the microphones ($\gamma^2$) indicates the degree of linear relationship between the signals~\cite{Roger2017}. It is a normalized quantity computed from the auto- ($P_{i,i}(f)$) and cross-power ($P_{i,j}(f)$) spectral density of the two microphone signals, defined as \cite{Roger2017}:
\begin{equation}\label{eq:gamma_exp}
    \gamma^2(f) = \frac{|P_{i,j}|^2}{P_{i,i}(f)P_{j,j}(f)}\,.
\end{equation}
The coherence varies from 0 to 1, with 0 meaning that there is no relationship between the signals. The coherence decreases with increasing distance between the microphones. The distance between the microphones plays a role in the coherence values. For this study, the coherence presented in the results section was computed for a pair of microphones spaced by $\eta_y=3$mm at $x/c=0.9275$.

Brooks~\cite{BROOKS1981} proposed the following exponential decay to model the coherence:
\begin{equation} \label{eq:gamma}
    \gamma^2(f) = \exp \left( -\frac{2 \pi f \eta_y}{U_c b_c} \right) \,,
\end{equation}
where $U_c$ is the convection velocity and $b_c$ is the Corcos' constant. The convection velocity is determined from the phase gradient spectrum ($\ud\phi_{i,j}/\ud f$) between two microphones in the streamwise direction:
\begin{equation}\label{eq:conv_vel}
    U_c = \frac{2 \pi \eta_y}{\ud\phi_{i,j}/\ud f} \,.
\end{equation}
The Corcos' constant $b_c$ is determined by fitting the experimental coherence given by Eq.~\ref{eq:gamma_exp} with the coherence model given by Eq.~\ref{eq:gamma}. The results section discusses the experimental coherence and the Corcos' constant determined for different roughness heights.


\subsection{Far-field noise measurements}
The trailing edge far-field noise was measured for each surface roughness element tested. A GRAS~40PH free-field microphone was placed at $z=1.5$~m perpendicular to the $x-y$ plane aligned with the TE at approximately the midspan location. Data was acquired at a sampling frequency of approximately 65.5~kHz ($2^{16}$~Hz) using PXIe-4499 Sound and Vibration modules embedded in a NI~PXIe-1073 chassis. The acquisition time was 30 seconds. The PSD of the far-field noise is calculated based on Welch’s method. The data was averaged using blocks of $2^{13} = 8192$ (0.125~s) samples and windowed using a Hanning windowing function with 50\% overlap, resulting in a frequency resolution of 8~Hz. The frequency was nondimensionalized by the boundary layer parameters at the trailing edge as the Strouhal number (Eq.~\ref{eq:St}).

\section{Results and Discussion} \label{sec:Results}
First, the wall pressure results are analyzed. The WPF was measured by 6 microphones. However, we discuss the data of one microphone, i.e., at midspan and at $x/c=0.9275$, for sake of brevity. The results observed for the other microphones are similar due to the statistical homogeneity of the flow at this location. Afterwards, the spanwise coherence between two microphones at $x/c=0.9275$ spaced by $\eta_y=3$mm is discussed for brevity. Finally, the far-field noise is analyzed for each case studied.

\subsection{Wall pressure fluctuations}

\begin{figure}[hbt!]
	\centering
    \begin{subfigure}[c]{0.49\textwidth}
        \centering
		\includegraphics[width=\textwidth]{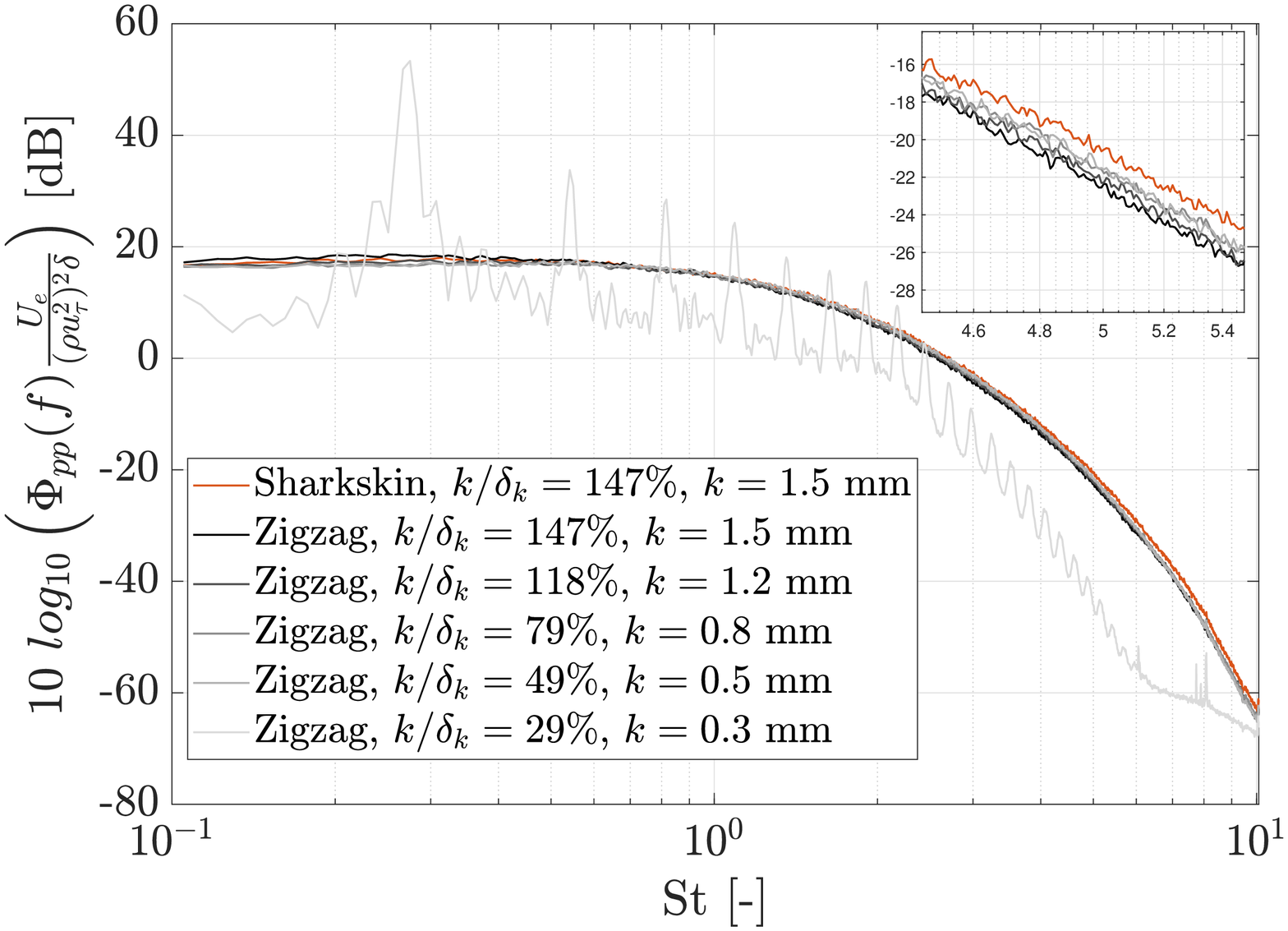}
		\caption{10 m/s.}
		\label{subfig:USP_10}
    \end{subfigure}
	\hfill
	\begin{subfigure}[c]{0.49\textwidth}
        \centering
        \includegraphics[width=\textwidth]{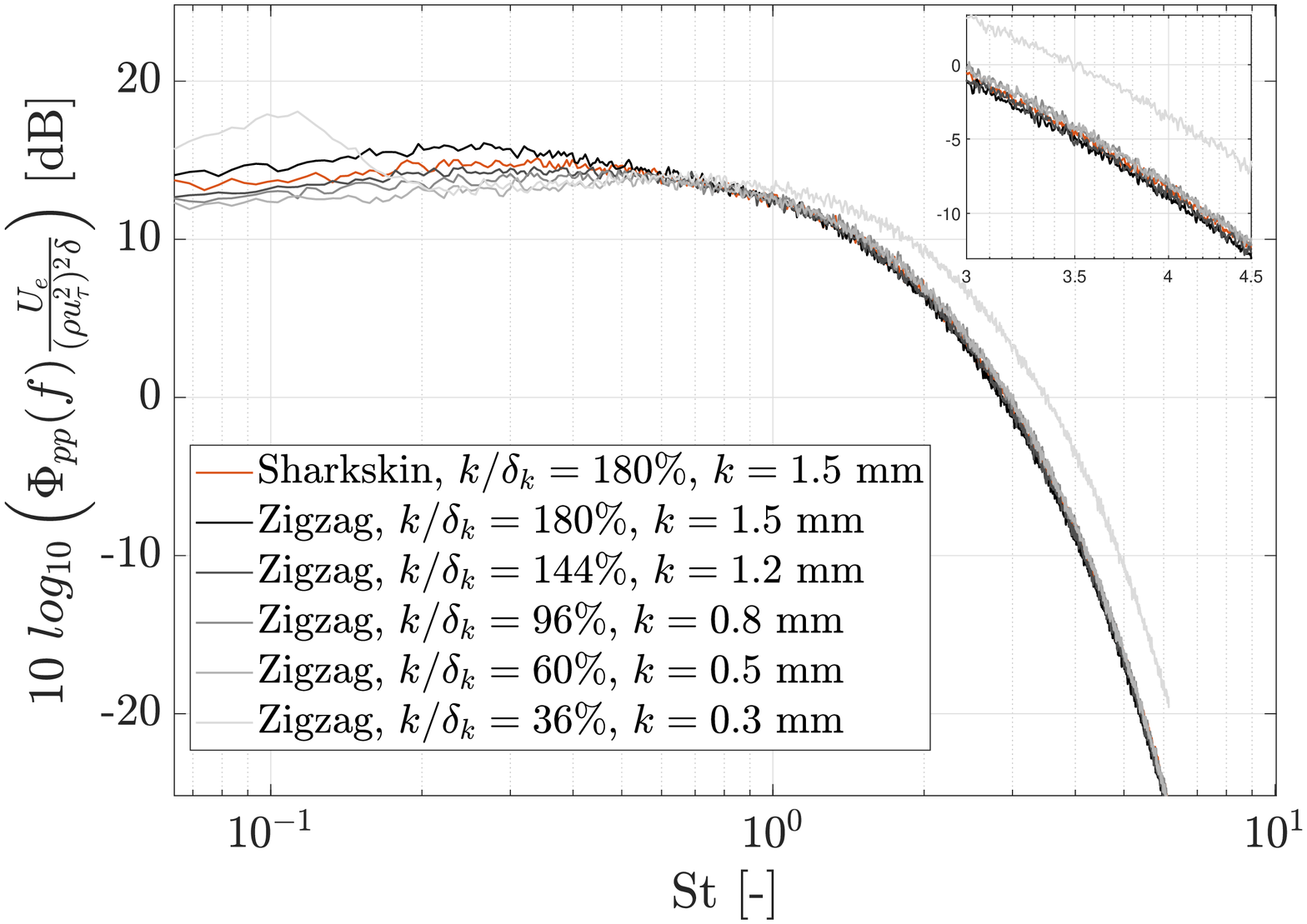}
        \caption{15 m/s.}
        \label{subfig:USP_15}
    \end{subfigure}
    \hfill
	\begin{subfigure}[c]{0.49\textwidth}
        \centering
        \includegraphics[width=\textwidth]{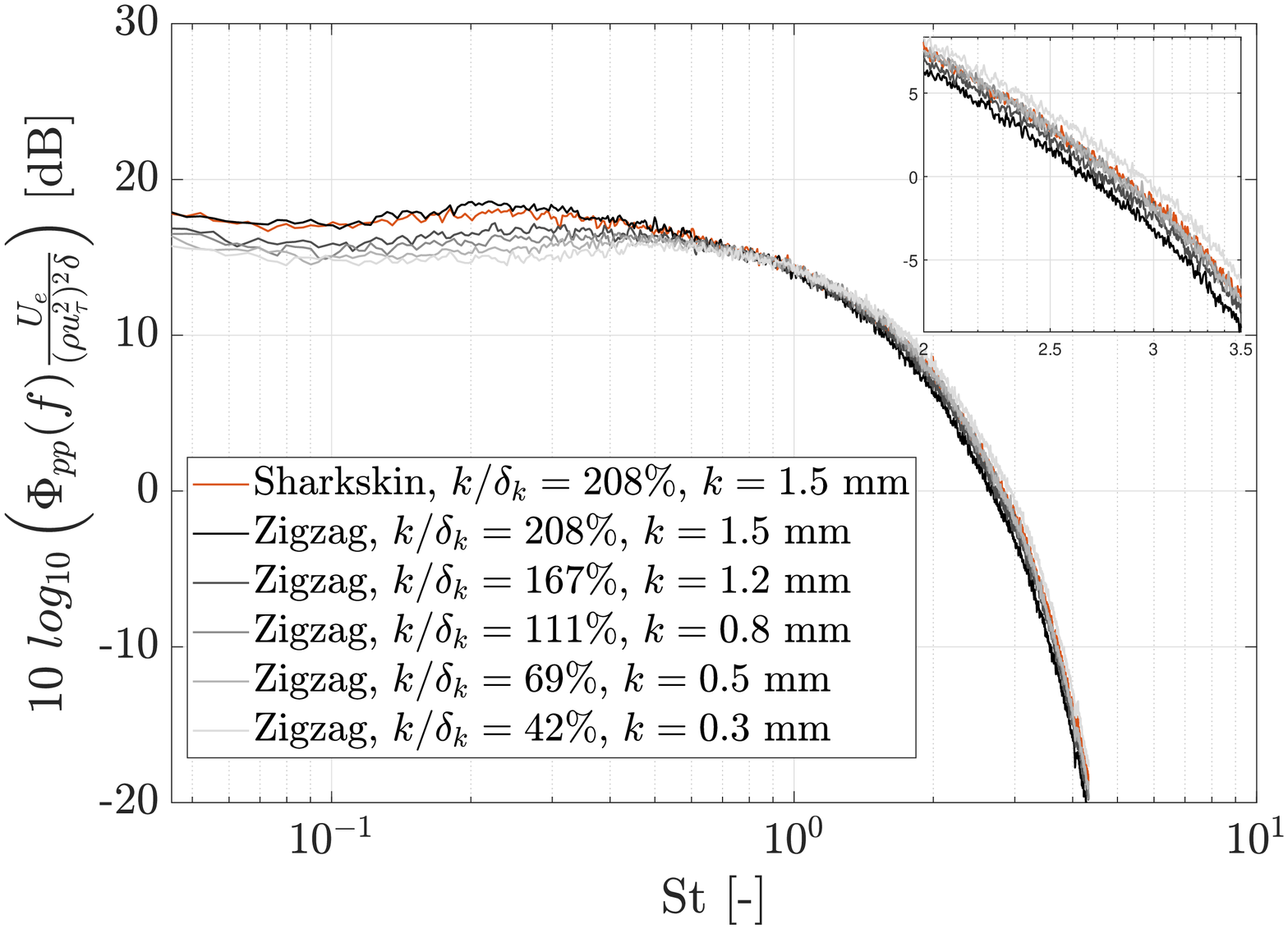}
        \caption{20 m/s.}
        \label{subfig:USP_20}
    \end{subfigure}
    \hfill
	\begin{subfigure}[c]{0.49\textwidth}
        \centering
        \includegraphics[width=\textwidth]{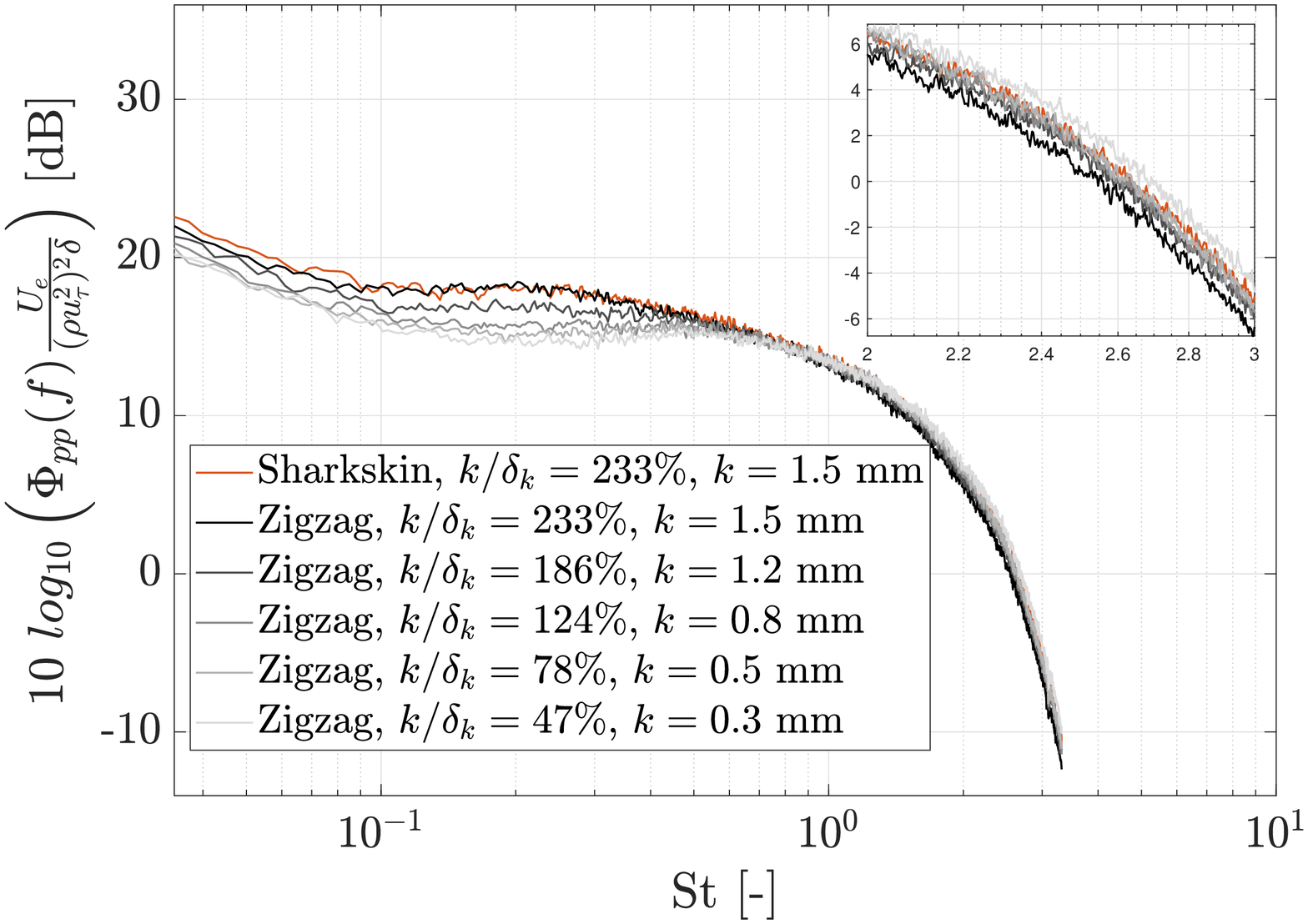}
        \caption{25 m/s.}
        \label{subfig:USP_25}
    \end{subfigure}
	\caption{Power spectral density of the wall pressure fluctuations at \boldmath{$x/c=0.9275$} at midspan for different surface roughness installed at \boldmath{$x/c=0.2$} for 10, 15, 20, and 25 m/s. The level is nondimensionalized by the boundary layer parameters at the trailing edge. Strouhal number based on the boundary layer thickness and edge velocity at the trailing edge. The insert zooms in on the high frequency range, showing more clearly the deviation of the spectrum for different trip heights.}
	\label{fig:USP}
\end{figure}

Figure~\ref{fig:USP} depicts the PSD of the wall pressure fluctuations at the trailing edge for the zigzag strips and sharkskin-like trip considering different heights and free-stream velocities.

\begin{figure}[b!]
	\centering
	\includegraphics[width=0.49\textwidth]{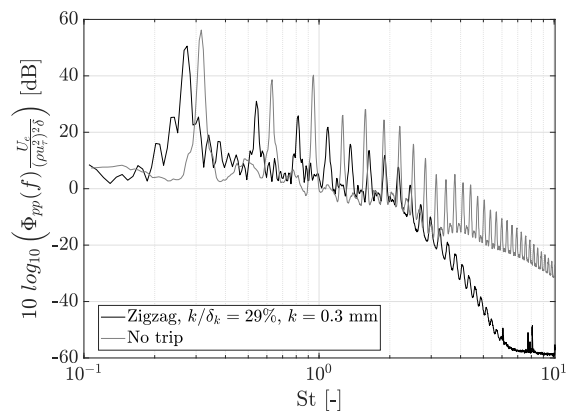}
	\caption{Power spectral density of the wall pressure fluctuations at \boldmath{$x/c=0.9275$} at midspan for non-tripped and tripped boundary layer for 10~m/s. The trip was installed at \boldmath{$x/c=0.2$} with a height of $k=0.3$~mm ($k/\delta_k=29\%$). The level is nondimensionalized by the boundary layer parameters at the trailing edge. Strouhal number based on the boundary layer thickness and edge velocity at the trailing edge.}
	\label{fig:USP_nontrip}
\end{figure}

For a trip height of 29\% $\delta_k$ the turbulent boundary layer is not fully developed, as inferred from Figs.~\ref{subfig:USP_10}~and~\ref{fig:USP_nontrip} from a tone centered at $St=0.27$ (270~Hz) with side lobes and presenting a harmonic behavior. A similar behavior is observed in literature~\cite{Yakhina2015,desquesnes2007,Chong2012}. Tones are usually encountered in low to moderate Reynolds number flows where the turbulent boundary layer is not fully developed at the TE~\cite{Arcondoulis2010}. According to Yakhina et al.~\cite{Yakhina2015} based on previous studies, Tollmien-Schlichting (T-S) waves develop as a first stage in a natural transition from laminar to turbulent flow. If the distance from the onset of T-S waves and the TE is large enough and/or the disturbances are strong enough, the boundary layer becomes turbulent at the TE. In this case the turbulence is scattered producing broadband TE noise. Otherwise, when the instability waves do not contribute to form a fully turbulent boundary layer at the TE, the instability waves develop only for certain frequency ranges with distinct amplification rates. Thus, the primary sound radiation is a broad spectral hump. In certain conditions the upstream propagating sound waves force the instabilities near the T-S waves onset, forming an acoustic feedback mechanism. In this case, a self-sustained motion establishes, leading to the emergence of a tone with higher harmonics. According to Desquesnes et al.~\cite{desquesnes2007} the tones can be attributed to a main and a secondary feedback loop. The main feedback loop occurs between the boundary layer on the pressure side and the acoustic source due to the interaction of the T-S waves with the trailing edge and causes the main tone observed. The secondary feedback loop occurs between the instabilities on the suction side and the TE.

The observations of Yakhina et al.~\cite{Yakhina2015} and Desquesnes et al.~\cite{desquesnes2007} show that the tripping device with $k/\delta_k=29\%$ introduced instabilities in the boundary layer that did not turn into turbulence up to the TE. For a non-tripped boundary layer, see Fig.~\ref{fig:USP_nontrip}, tones are observed at harmonics of the fundamental frequency without side lobes which are likely due to the feedback loop between the T-S waves developed naturally and the TE. For a tripped boundary layer, a tone is observed with side lobes with the spectrum presenting a broadband behavior which resembles the one for a fully developed turbulent flow. The tones indicate that the feedback loop between the T-S waves and the TE is taking place, generating the center tone but with other instabilities also introducing tones of smaller intensity. These observations might be a consequence of:
\begin{enumerate}
    \item the instabilities introduced by the roughness differ from the instabilities of a undisturbed boundary layer;
    \item the trip changed the boundary layer parameters which is not captured by the values used to nondimensionalize the spectrum since they were obtained from XFOIL;
    \item the instabilities were more developed at the TE than for the case without roughness since the roughness has introduced strong disturbances in the boundary layer, leading to the side lobes.
\end{enumerate}
The real phenomenology might be one or a combination of these hypotheses, but to verify these hypotheses, boundary layer measurements would be required and it will be the next step for this research. Furthermore, the boundary layer starts to slightly present the behavior of a fully developed turbulent flow (broadband behavior) probably because the instabilities were more developed at the TE than for the case without roughness.

As can be inferred from Fig.~\ref{subfig:USP_15}, a roughness height of 36\% $\delta_k$ tested at 15~m/s results in a spectrum with considerable discrepancy from the other cases (up to 5~dB) for $St>1.3$. Besides, a hump is also observed for $St<0.15$ indicating that larger turbulence scales are present in the flow with a higher energy level than the other roughness heights. This roughness height probably did not cause a transition just after the trip, resulting in turbulent structures that were not fully developed at the TE. These structures cause the hump in the spectrum for low frequencies and the higher spectrum level for high frequencies.

For all velocities the WPF level increased with increasing the roughness elements height for $St<0.7$, being more evident as the ratio $k/\delta_k>100\%$, reaching differences up to 3.5~dB, see Fig.~\ref{fig:USP}. For roughness elements smaller than the boundary layer thickness the difference between the spectra was smaller than 0.5~dB, but as soon as the roughness is higher than the boundary layer thickness the differences increase considerably. This increase for low frequencies is probably because of the increase in boundary layer thickness as the roughness height increased. The development of a thicker boundary layer due to roughness was observed in other studies~\cite{Fernanda2020}. A thicker boundary layer results in larger eddies that contain more energy, culminating in a higher energy level for low frequencies as observed in this study. This is in agreement with the energy cascade concept since it states that the large-scale motions are strongly influenced by the flow conditions~\cite{Pope}. The boundary layer thickness $\delta$ utilized to nondimensionalyze the spectra in Fig.~\ref{fig:USP} was determined from XFOIL, which considers that the transition happens at $x/c=0.2$, resulting in the same boundary layer thickness regardless of the trip height applied. Hence, the nondimensionalization uses the correct variables since for low frequencies the scaling should be based on the boundary layer outer parameters, e.g., boundary layer thickness~\cite{Devenport}; however, it does not use the actual boundary layer parameter since this parameter was not measured. To confirm this conjecture about the boundary layer thickness, boundary layer measurements are required which will be the next step for this research. If we consider that the boundary layer becomes thicker for higher roughness, the nondimensional spectrum level decreases for higher roughness, giving a good indication that this conjecture is valid.

An overlap of the spectra for $0.7<St<1$ was observed independently of the roughness height tested, see Fig.~\ref{fig:USP}. According to the energy cascade theory and the Kolmogorov hypotheses, the outer flow parameters affect only the large scales and, as the energy is transferred from the large to the small scales, all the information about the geometry of the large scales is lost~\cite{Pope}. This suggests that the mid-frequency range is not affected by the mean flow. Furthermore, the mid-frequency range is believed to be scaled by the outer and inner boundary layer parameters, e.g., viscosity and wall shear stress~\cite{Devenport}. Hence, the overlap of the spectra for $0.7<St<1$ shows that these turbulent eddies are no longer affected by the large and small scales, and consequently, are also not affected by the surface roughness.

For $1<St<6$ an interesting behavior is observed: for the range of approximately $50\%<k/\delta_k<110\%$ the curves agree very well. However, for $35\%<k/\delta_k<50\%$ the spectrum level increases and for $k/\delta_k>110\%$ it decreases in comparison with the spectrum curves for $50\%<k/\delta_k<110\%$. To explain this behavior, two facts should be highlighted: 1. according to the energy cascade concept, the small eddies are no longer affected by the large eddy characteristics, but instead have a universal behavior~\cite{Pope}; and 2. they are located near the wall. Therefore, their scaling is determined by the inner boundary layer parameters~\cite{Devenport}. For all roughness heights that resulted in a turbulent boundary layer at the TE, the behavior of the spectra for $1<St<6$ is similar but with different levels, which shows the universal behavior of these scales. For the difference in levels, this is most probably associated with the different development of the boundary layer as the roughness height increases. For $50\%<k/\delta_k<110\%$, it is believed that the roughness is introducing vortical structures in the boundary layer (as observed by~\cite{Elsinga2012}) that are strong enough to trigger turbulent spots that will mix and cause a boundary layer transition close to the roughness element. Because the transition is occurring in similar positions, it is believed that the inner boundary layer for these cases are similar. For $35\%<k/\delta_k<50\%$, the roughness probably introduces vortices and instabilities in the boundary layer but not strong enough to cause a transition just after the roughness. Therefore, the turbulent boundary layer developed is different from the one where the transition happened just after the roughness. For $k/\delta_k>110\%$, the roughness considerably changed the inner and outer part of the boundary layer since the trip is blocking the boundary layer entirely. The transition occurs just after the roughness, but because of the blocking of the entire boundary layer by the roughness a different boundary layer is developed that will have different inner parameters when compared to the $50\%<k/\delta_k<110\%$ case.

Regarding the sharkskin-like geometry, this roughness seems to be an effective way of generating a turbulent boundary layer. The sharkskin trip behaves as a zigzag strip of the same height for low frequencies ($St<0.7$), producing a similar hump. This indicates that the sharkskin disturbs the large turbulent scales in the same way as the zigzag strip. For high frequencies ($St>1$), the sharkskin behaves as a trip with a height $50\%<k/\delta_k<110\%$. This might be associated to its open area when compared to the zigzag strip. Contrary to the zigzag strip, the sharkskin geometry does not block the boundary layer completely, leading to the development of a inner boundary layer that resembles one  from a zigzag of a smaller height.

\subsection{Spanwise coherence}
\begin{figure}[b!]
	\centering
    \begin{subfigure}[c]{0.49\textwidth}
        \centering
		\includegraphics[width=\textwidth]{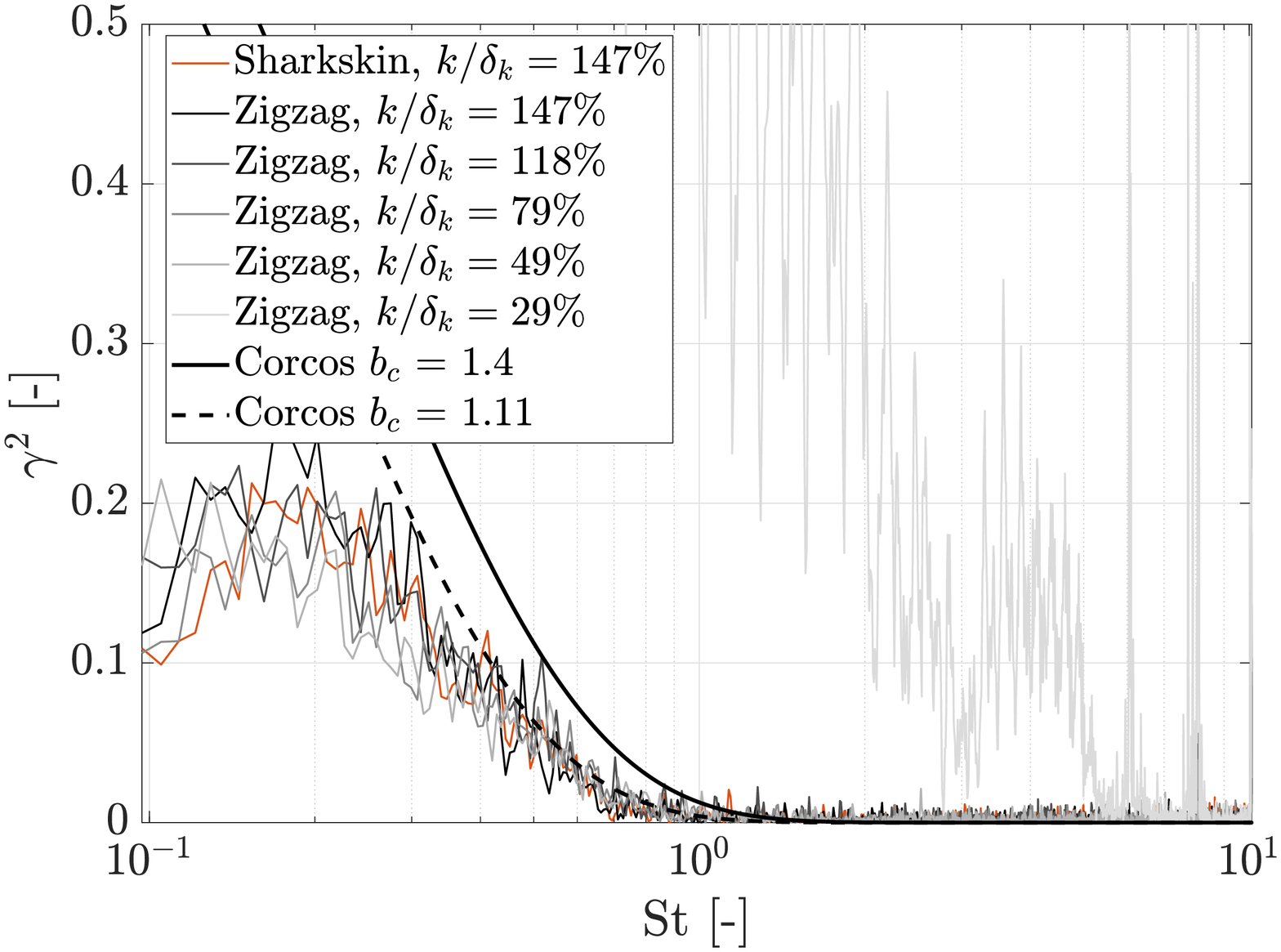}
		\caption{10~m/s.}
		\label{subfig:coherence_10}
    \end{subfigure}
	\hfill
	\begin{subfigure}[c]{0.49\textwidth}
        \centering
        \includegraphics[width=\textwidth]{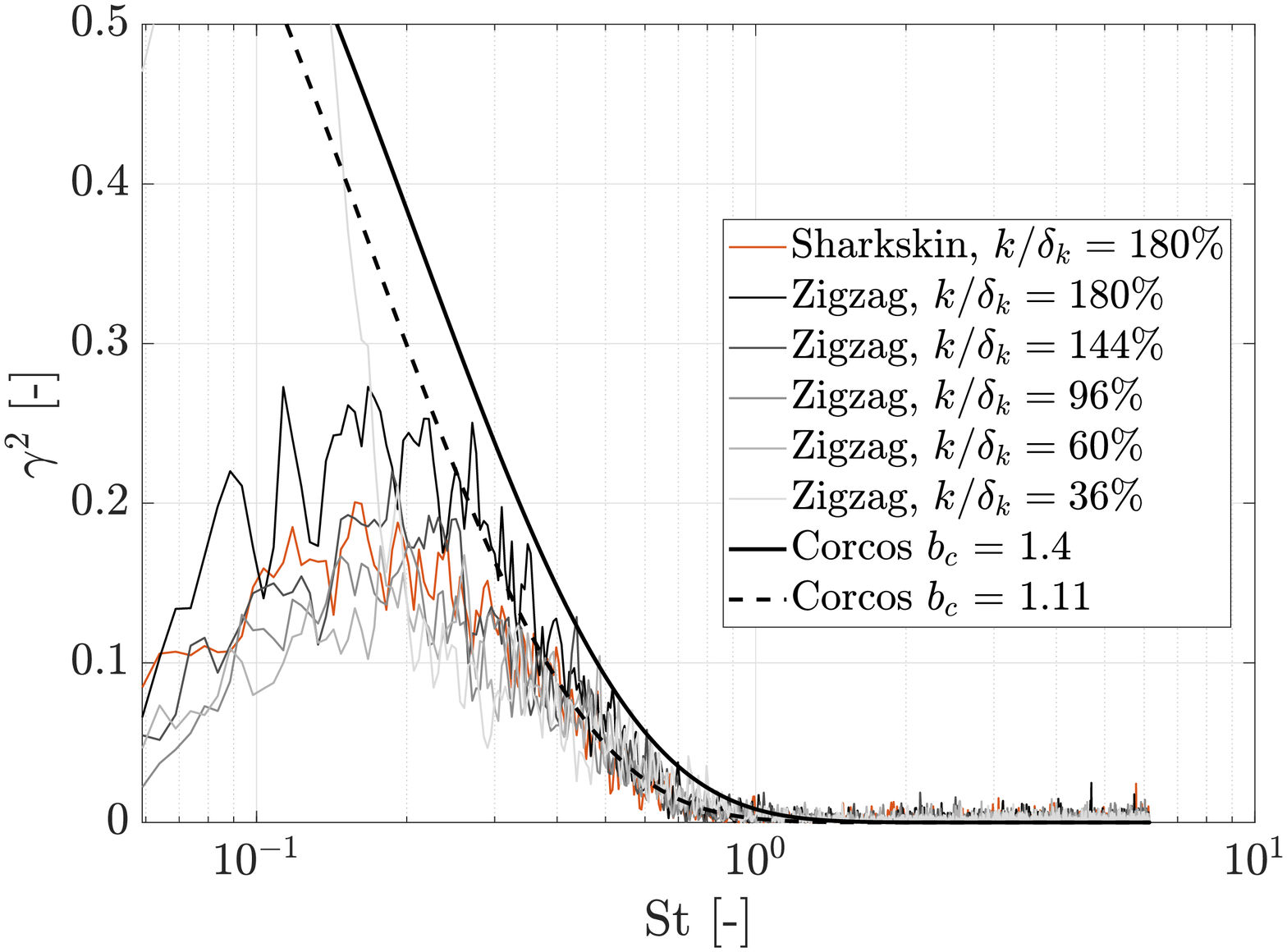}
        \caption{15~m/s.}
        \label{subfig:coherence_15}
    \end{subfigure}
    \hfill
	\begin{subfigure}[c]{0.49\textwidth}
        \centering
        \includegraphics[width=\textwidth]{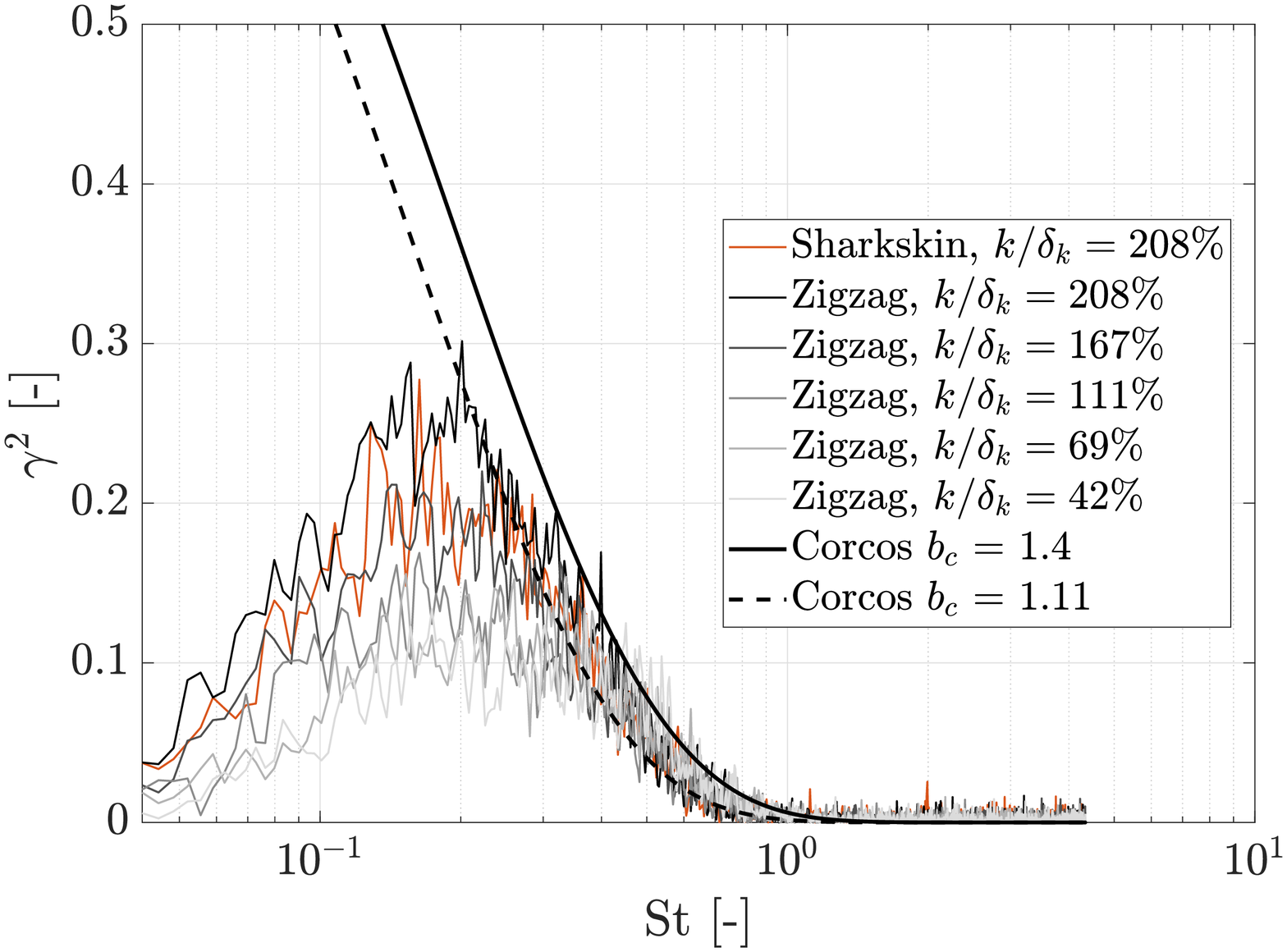}
        \caption{20~m/s.}
        \label{subfig:coherence_20}
    \end{subfigure}
    \hfill
	\begin{subfigure}[c]{0.49\textwidth}
        \centering
        \includegraphics[width=\textwidth]{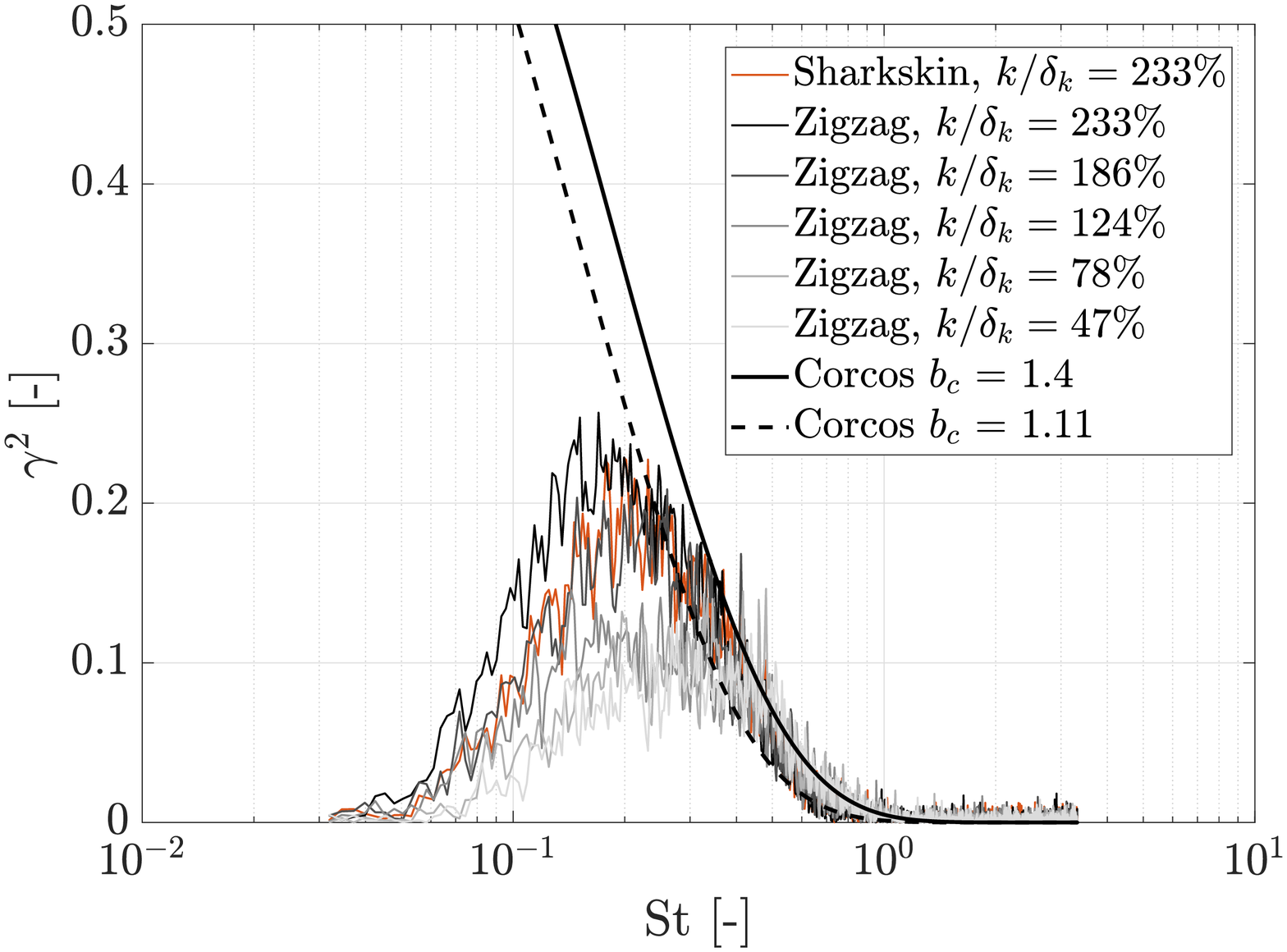}
        \caption{25~m/s.}
        \label{subfig:coherence_25}
    \end{subfigure}
	\caption{Coherence between two microphones in the spanwise direction spaced by \boldmath{$\eta_y=3$}~mm at \boldmath{$x/c=0.9275$} for different surface roughness installed at \boldmath{$x/c=0.2$} for 10, 15, 20, and 25~m/s. Strouhal number based on the boundary layer thickness and edge velocity at the trailing edge. Exponential decay given by Eq.~\ref{eq:gamma} for two values of the Corcos' constant: \boldmath{$b_c=1.4$} from the literature ~\cite{STALNOV201650}, and \boldmath{$b_c=1.1$} from the fitting of Eq.~\ref{eq:gamma} with the experimental data.}
	\label{fig:coherence}
\end{figure}

Figure~\ref{fig:coherence} shows the spanwise coherence for different trip heights and velocities. In this figure, Corcos' spanwise coherence, given by Eq.~\ref{eq:gamma}, is also plotted for $b_c=1.4$~\cite{STALNOV201650} and for $b_c=1.1$ that is the average value for the Corcos' constant considering all trip heights and velocities. The convection velocity was determined by the phase gradient spectrum between two microphones in the streamwise direction, as shown in Eq.~\ref{eq:conv_vel}. The calculated average convection velocity for all cases studied was $U_c/U=0.61$. This value is used in Eq.~\ref{eq:gamma}.

For the trip height $k/\delta_k=29\%$ (Fig.~\ref{subfig:coherence_10}) and $k/\delta_k=36\%$ (Fig.~\ref{subfig:coherence_15}), we observe that the coherence does not follow the same behavior as for the other trips. For $k/\delta_k=29\%$, the coherence is extremely high in the entire frequency range, indicating that large correlated structures are presented in the flow in the spanwise direction, which agrees with the observations made regarding the WPF spectra for this trip. Hence, this strongly suggests that the boundary layer is not fully turbulent at the TE. For $k/\delta_k=36\%$, it is observed that the coherence value follows the one for the other trips for $St>2$; however, it increases greatly for low frequencies. This indicates that a strongly coherent structure is still present in the boundary layer, which is no longer observed for higher trip heights. Therefore, these observations confirm the idea that the roughness heights where $k/\delta_k<36\%$ do not develop a fully turbulent boundary layer at the TE.

The spanwise coherence presents roughly the same decay for all trip heights for a specific velocity, as can be inferred from Fig.~\ref{fig:coherence}, but with higher coherence value for low frequencies as the trip height increases. This indicates that stronger coherent structures are formed in the spanwise direction for higher roughness.

The sharkskin-like trip has coherence values similar to smaller zigzag strips, see Fig.~\ref{fig:coherence}. Due to its non-blocking effect, the sharkskin trip probably induces less strongly coherent structures in the spanwise direction, having an effective height that is smaller than the zigzag strip.

The average Corcos' constant for all the measurements performed ($b_c=1.1$) is smaller than the one from the literature ($b_c=1.4$~\cite{STALNOV201650}). The coherence curves for 25~m/s are the ones that approximate the most to the Corcos' constant of 1.4. Furthermore, the coherence given by Eq.~\ref{eq:gamma} matches well with the decay of the experimental coherence for high frequencies. The same behavior of the experimental coherence has also been observed in the literature~\cite{STALNOV201650,Sanders2018}.

\subsection{Far-field noise}
\begin{figure}[h!]
	\centering
    \begin{subfigure}[c]{0.49\textwidth}
        \centering
		\includegraphics[width=\textwidth]{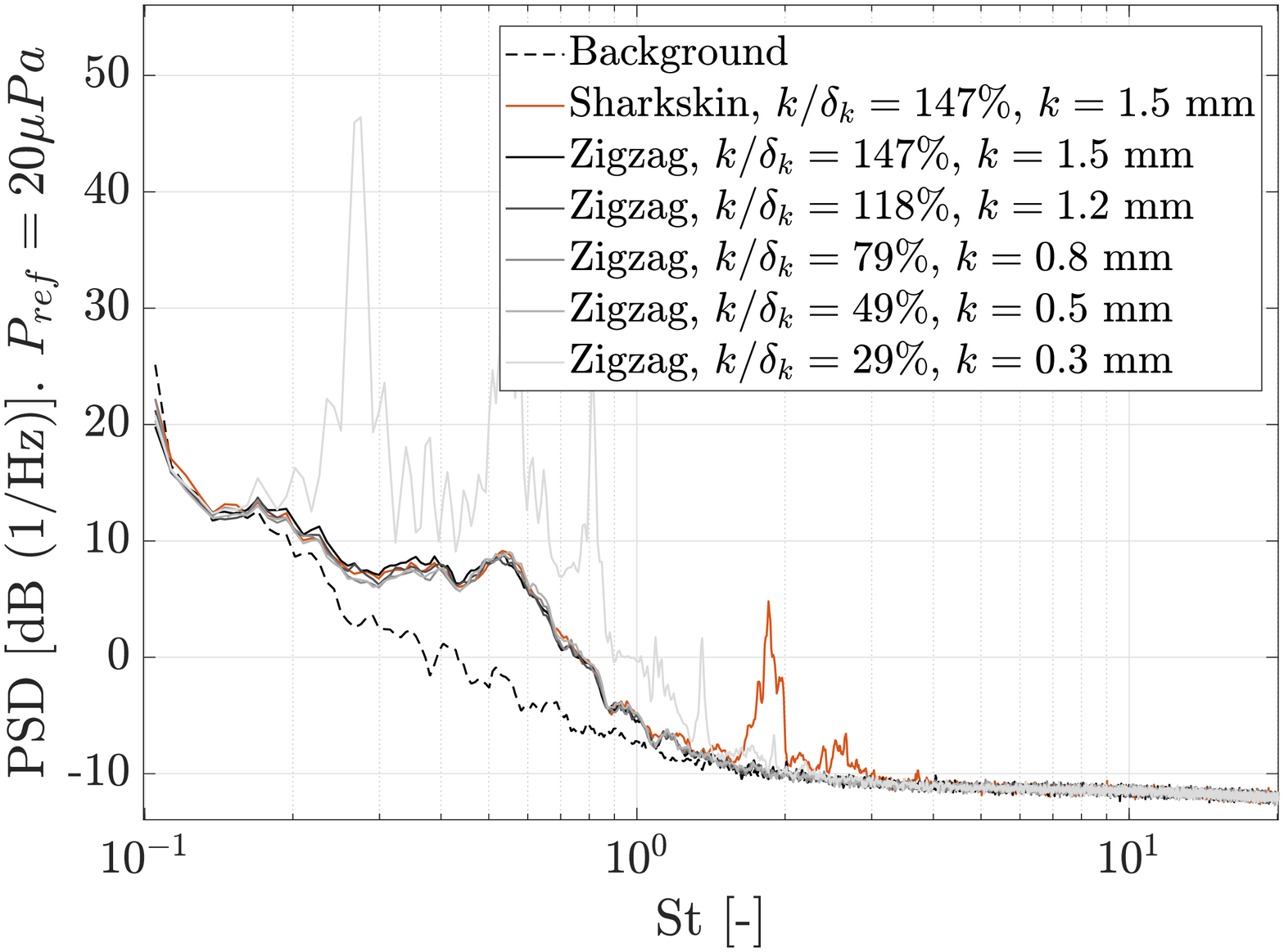}
		\caption{10 m/s.}
		\label{subfig:FFM_10}
    \end{subfigure}
	\hfill
	\begin{subfigure}[c]{0.49\textwidth}
        \centering
        \includegraphics[width=\textwidth]{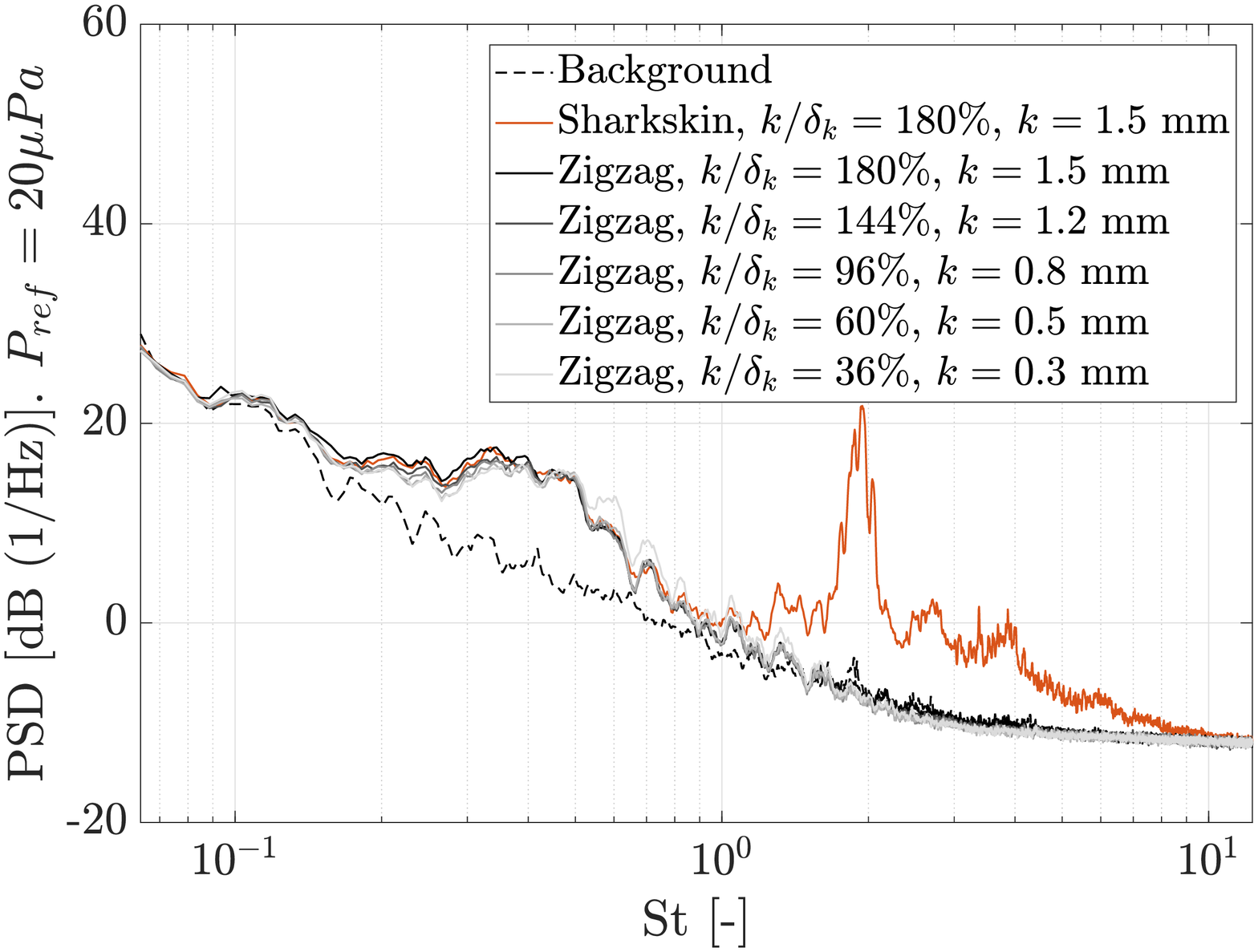}
        \caption{15 m/s.}
        \label{subfig:FFM_15}
    \end{subfigure}
    \hfill
	\begin{subfigure}[c]{0.49\textwidth}
        \centering
        \includegraphics[width=\textwidth]{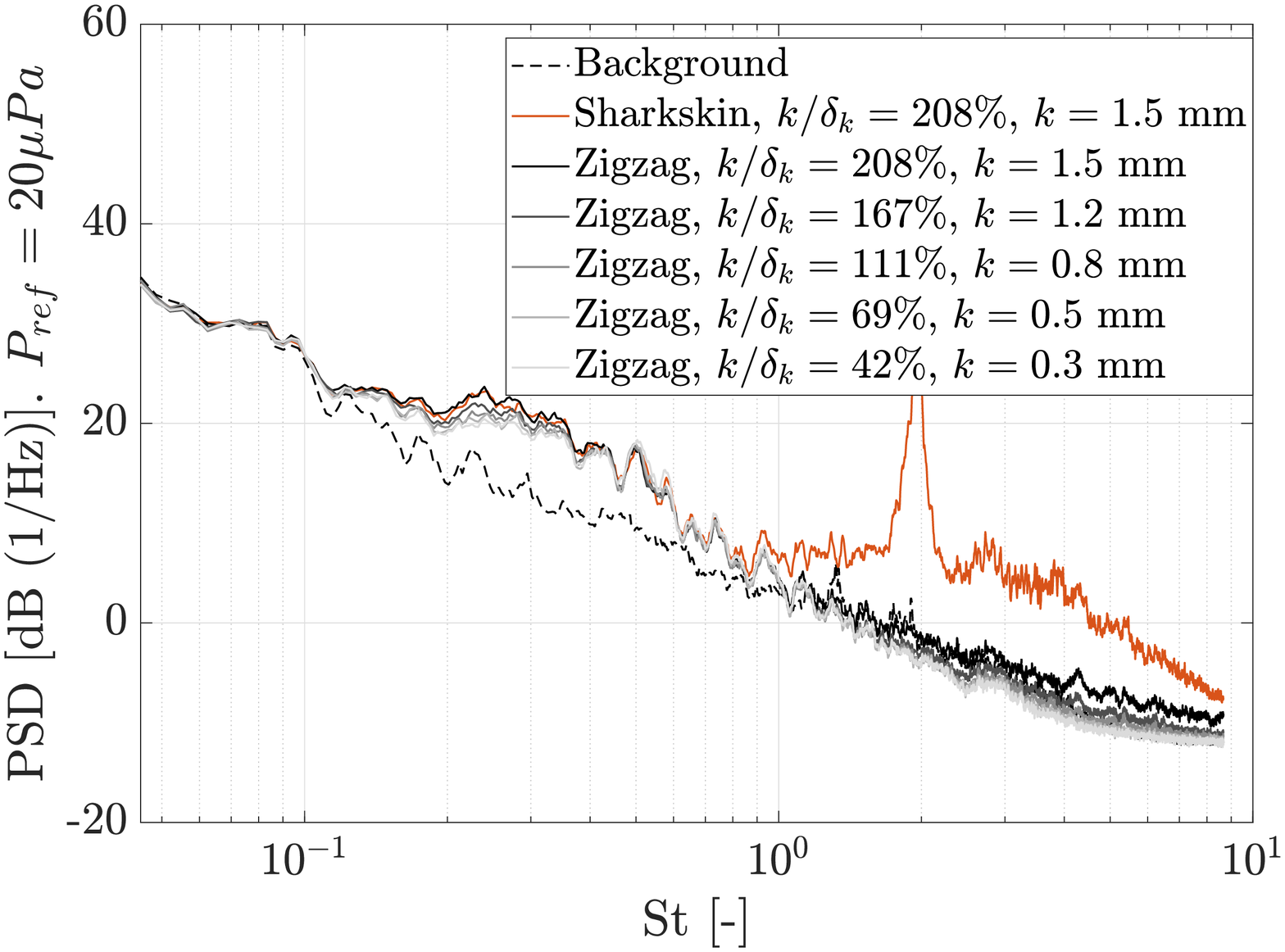}
        \caption{20 m/s.}
        \label{subfig:FFM_20}
    \end{subfigure}
    \hfill
	\begin{subfigure}[c]{0.49\textwidth}
        \centering
        \includegraphics[width=\textwidth]{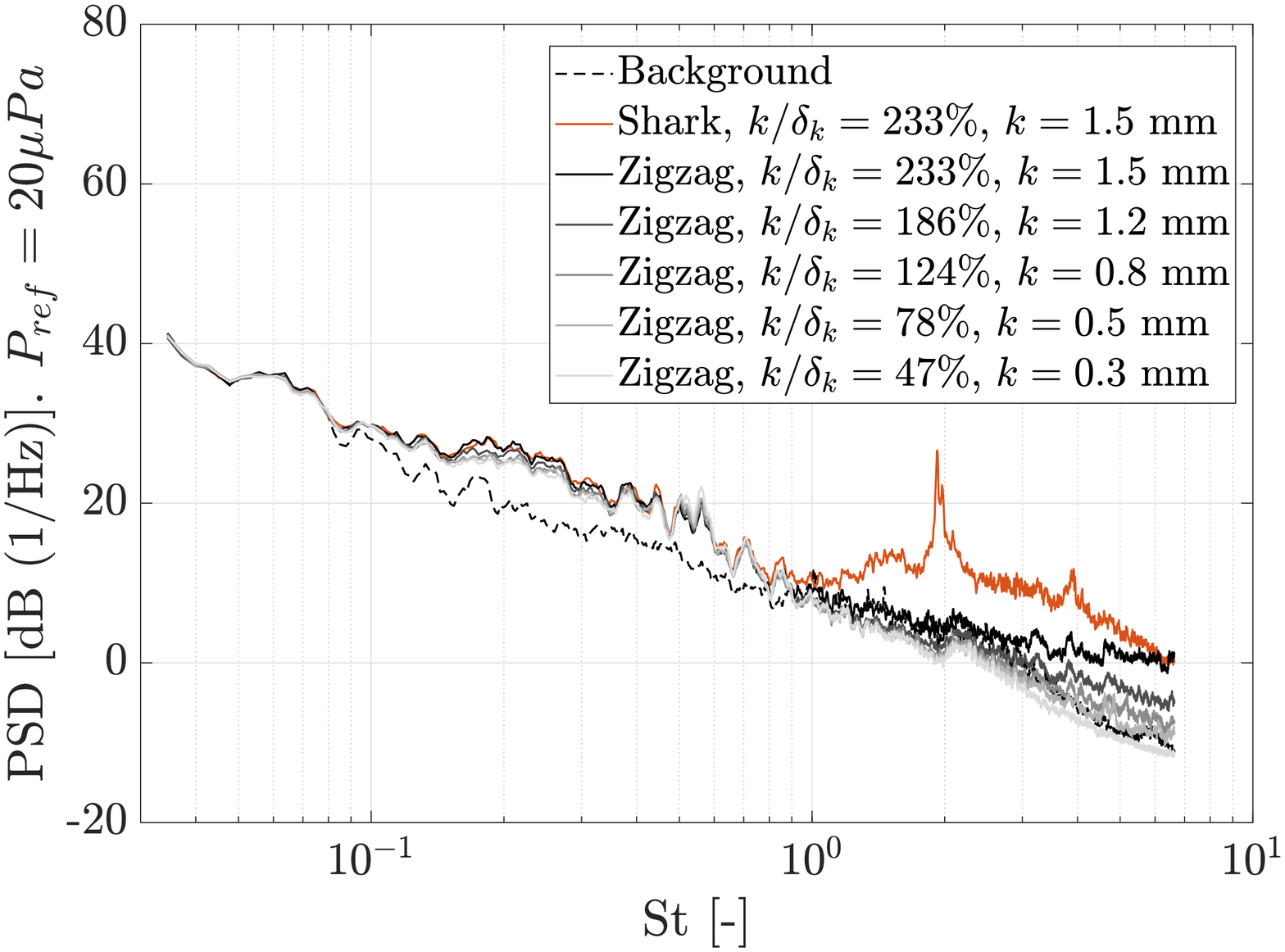}
        \caption{25 m/s.}
        \label{subfig:FFM_25}
    \end{subfigure}
	\caption{Power spectral density of the far-field noise for a microphone at \boldmath{$z=1.5$}~m perpendicular to the trailing edge for different roughness installed at \boldmath{$x/c=0.2$} for 10, 15, 20, and 25~m/s. Strouhal number based on the boundary layer thickness and edge velocity at the trailing edge.}
	\label{fig:FFM}
\end{figure}

The far-field noise was measured for different flow velocities and different roughness element heights, as depicted in Fig.~\ref{fig:FFM}. The background noise of the wind tunnel is also included in the figure.

For the trip height $k/\delta_k=29\%$ (Fig.~\ref{subfig:FFM_10}), we observe a far-field noise with the same tones present in the WPF spectrum. Amiet's trailing edge noise model relates the far-field noise directly to the wall pressure fluctuations at the TE~\cite{Amiet1976}. Hence, the feedback loop between the T-S waves and the TE is the noise source for the tones observed in the far-field spectrum. However, for $k/\delta_k=36\%$ (Fig.~\ref{subfig:FFM_15}), an increase in the far-field noise is observed for $0.5<St<1.3$, which is not observed in the WPF spectrum. Therefore, the increased far-field noise for this roughness height is not due to the WPF at the TE but instead it is due to an unknown noise source that was observed only in this measurement. Furthermore, the hump noticed in the WPF for low frequencies is no longer present in the far-field noise. This discrepancy might be due to the further development of the boundary layer since the WPF was measured at $x/c=0.9275$ and not exactly at the TE. This assumption requires further investigation.

The increase of the WPF spectrum level for low frequencies as the roughness height increased is also observed for the far-field noise spectrum, reaching deviations up 2~dB compared to the lowest roughness height. The far-field level increases for $St<0.4$, which is relatively close the the increased level observed for the WPF ($St<0.7$), showing that the far-field noise increase in this frequency range is most probably due to the unsteadiness reaching the airfoil trailing edge, which was affected by the roughness.

For extremely high roughness ($k/\delta_k>180\%$), an increase in the far-field noise for high frequencies is observed. This increased level does not occur in the WPF. Therefore, it is believed that the high-frequency noise is coming from the trip itself and not from its influence in the trailing edge noise. The increase of the far-field noise by the trip was also observed by Hutcheson and Brooks~\cite{Hutcheson}. The sharkskin trip generates an extremely high-frequency noise with a tone at $St \approx 2$, which is coming from the sharkskin trip itself and not from its effects in the TE wall pressure.

\section{Conclusions}
We investigated the influence of surface roughness element height and shape in the trailing edge near- and far-field noise generation. The surface roughness elements studied are a zigzag strip and a novel sharkskin-like geometry. The wall pressure fluctuations close to the TE and the far-field noise were measured. The near-field analysis of the roughness effects on the wall pressure fluctuations and spanwise coherence at the trailing edge are addressed for the first time in open literature to the knowledge of the authors. This study showed that the height of the surface roughness elements affects the wall pressure spectrum in the low- and high- frequency ranges as well as the far-field noise. These observations are also applicable to pieces of equipment that have a rough surface due to erosion. Therefore, determining the correct geometric characteristics of the roughness elements and surface roughness is essential to obtain accurate experimental data and noise prediction.

For a roughness height $k/\delta_k<36\%$ the boundary layer was not fully turbulent at the TE. For $k/\delta_k<29\%$, a tone is observed with side lobes and their harmonics in the WPF spectrum with a broadband behavior resembling the one for a fully developed turbulent flow. It is believed that the tone is caused by the feedback loop between the T-S waves and the TE with other instabilities induced by the roughness also introducing tones of smaller intensity. The WPF spectrum also presents a broadband behavior being similar to a fully developed turbulent boundary layer spectrum. Tones are likewise observed in the far-field noise. A roughness height $k/\delta_k>42\%$ suppressed any T-S wave radiation and a fully turbulent boundary layer was established. Hence, the roughness element hastens the transition compared to a naturally developed boundary layer but for $k/\delta_k<36\%$ the instabilities were not fully developed at the TE.

As the surface roughness height increased, a hump in the WPF and the far-field noise spectrum is observed for low frequencies $St<0.7$. The level increase is believed to be a result of the development of a thicker boundary layer as the trip height increases. Boundary layer measurements were not performed in this study, with the nondimensionalization of the spectrum done by parameters from XFOIL, using a prescribed location of transition at $x/c=0.2$, resulting in the same boundary layer thickness regardless of the roughness applied. Thus, to confirm the influence of the roughness for the low frequency range and its scaling, boundary layer measurements will be performed as a next step for this research.

The roughness height and geometry do not affect the WPF and far-field noise spectra, and consequently, the turbulent eddies in the mid-frequency range $0.7<St<1$. This is supported by the energy cascade theory and the Kolmogorov hypotheses since they stated that the outer flow affects only the large scales (low-frequency) and, as the energy is transferred from the large to the small scales, all the information about the geometry of the large scales is lost~\cite{Pope}.

In the high-frequency range ($1<St<6$), the roughness height affects the WPF spectra indicating that the roughness alters the inner boundary layer parameters. The WPF spectrum has the same curve behavior for all roughness heights tested but it presents different levels depending on the roughness height, which is in accordance with the universal behavior of the small scales stated in the energy cascade theory~\cite{Pope}. However, the roughness height affects the level of the WPF spectra, indicating that the roughness does influence the inner boundary layer characteristics since the high-frequency scaling is determined by the inner boundary layer parameters~\cite{Devenport}.

Based on the results of this study, trip heights in the range $50\%<k/\delta_k<110\%$ are strongly recommended for wind tunnel tests because:
\begin{enumerate}
    \item they cause a minimum level increase in the low-frequency range for the TE wall pressure spectrum and the far-field noise;
    \item they generate similar WPF spectrum for high-frequencies;
    \item they hasten the transition and effectively suppress tones in the spectrum considered to be due to the feedback loop between T-S waves and TE.
\end{enumerate}
These findings are consistent with literature where its was reported that roughness elements with height in the range $50\%<k/\delta_k<110\%$ trigger transition just after the trip and develop a boundary layer that is self-similar. This hypothesis is going to be verified in follow-up tests.

Regarding the sharkskin-like geometry, this trip is an effective way of generating a turbulent boundary layer; however, it results in high levels of the far-field noise in the high-frequency range. Therefore, this geometry is not recommended for wind tunnel tests.

\section*{Funding Sources}

Part of this research received financial support from the European Commision through the H2020-MSCA-ITN-209 project zEPHYR (grant agreement No 860101). 

\section*{Acknowledgments}
The authors are grateful to Ing. W. Lette, ir. E. Leusink, S. Wanrooij amongst others for their assistance during the wind tunnel tests. The authors would like to thank TNO and the Maritime Research Institute Netherlands (MARIN), particularly Dr.~Johan Bosschers, Dr.~Roel M\"{u}ller, and Dr.~Christ de Jong,  for the insightful discussions. The authors would also like to thank Dr. W. Nathan Alexander and Dr. N. Agastya Balantrapu for their efforts in preparing and sharing the dataset used for comparison in this research.

\bibliography{sample}

\end{document}